\documentclass[11pt]{article}
\usepackage{graphicx}
\author{}
\usepackage{enumerate}
\usepackage{multirow}
\usepackage{amsmath, amsthm, amssymb,framed}
\usepackage{epsfig}
\usepackage{setspace}
\usepackage{acronym}
\usepackage[linesnumbered,ruled,vlined]{algorithm2e}
\usepackage{comment}
\usepackage{relsize}
\usepackage{multicol}
\usepackage{adjustbox}
\usepackage{epstopdf}
\usepackage{float}
\usepackage{enumitem}

\usepackage[T1]{fontenc}
\usepackage{longtable}
\usepackage{lmodern}
\usepackage{hyperref}
\usepackage{xr}
\usepackage{xcolor}
\definecolor{blue}{RGB}{25,25,112}
\usepackage{subcaption} 
\usepackage{float}

\usepackage{hyperref}      
\usepackage{caption} 

\begin{document}
\title{\huge \color{blue} \textbf{Technical Supplement Report} \\ \Large ``\textbf{Full-Duplex FBMC/QAM MIMO Systems: Transceiver Design and Optimization}''\\ %IEEE Transactions on Wireless Communications Manuscript ID: TW-Nov-25-3135
}
\author{Sudhakar Rai, Prem Singh, Senior Member, IEEE, Ekant Sharma, Senior Member,\\ IEEE, Aditya K. Jagannatham, Senior Member, IEEE and Lajos Hanzo,\\ Life Fellow, IEEE}

\date{\empty}\maketitle 

This technical supplement is a companion document to the main
paper. All core theoretical derivations, SINR expressions (eqs.~(26) and~(34)),
SE analysis (eq.~(36)), and SE optimization Algorithm~1 are contained in the main paper.
This supplement provides additional context, figures, and comparisons developed
during the peer-review process that could not be accommodated within the main
paper's page limit.
\section{Notation Tables}
Table 1 and Table 2 define all notation used in the FBMC/QAM system model for uplink/downlink signal processing (in Section~II-A, Section~II-B, Section~II-D, Section~II-F), channel modeling (in Section~II-C and Section~II-E), SINR expressions (in (26) and (34)), and the proposed SE maximization problems (\textbf{P1}-\textbf{P6}). These unified tables supersede the
separate notation discussions in Section II of the main paper.
\begin{table}[H]
\centering
\footnotesize
\caption{Nomenclature (System Model and Channels)}
\label{tab_1}
\vspace{-10pt}
\renewcommand{\arraystretch}{1.1}
\begin{tabular}{ |p{3cm}||p{9.5cm}| }
\hline
\textbf{Notation} & \textbf{Description} \\
\hline

$\mathbf{d}_{j,b}^{s}(n)$ 
& Information symbol vector for the $b$-th subcarrier group at FBMC symbol index $n$, where $s \in \{\text{ul, dl}\}$ and $j \in \{k,t\}$. \\
\hline

$\mathbf{x}_{j,b}^{s}(n)$ 
& Filtered time-domain FBMC signal vector corresponding to $\mathbf{d}_{j,b}^{s}(n)$. \\
\hline

$\mathbf{s}_{j}^{s}(i)$ 
& Transmitted FBMC/QAM signal at discrete-time index $i$. \\
\hline

$\tilde{\mathbf{h}}_{r,k}$ 
& Uplink multi-tap channel impulse response from transmitter $k$ to receiver $r$. \\
\hline

$\tilde{\mathbf{H}}_{r,k}$ 
& Toeplitz channel matrix constructed from $\tilde{\mathbf{h}}_{r,k}$. \\
\hline

${\mathbf{H}}_{r,k}^{b,b'}(n)$ 
& Uplink effective FBMC channel matrix from subcarrier group $b$ to $b'$ at delay index $n$, including filtering and multipath effects. \\
\hline

$\sum\limits_{n=-L}^{L-1} \sum\limits_{b=0}^{B-1} {\mathbf{H}}_{r,k}^{b,b'}(n)$ 
&
\begin{minipage}[t]{9.3cm}
\vspace{2pt}
$\underbrace{
{\mathbf{H}}_{r,k}^{\text{desire},b'} + {\mathbf{H}}_{r,k}^{\text{ICI},b'}
}_{\text{at } b'}$
\vspace{2pt}

$+\;\underbrace{\sum\limits_{\substack{b=0 \\ b \ne b'}}^{B-1} {\mathbf{H}}_{r,k}^{b,b'}(0)}_{\text{inter-group interference}}$

\vspace{2pt}

$+\;\underbrace{\sum\limits_{\substack{n=-L \\ n \ne 0}}^{L-1}
\sum\limits_{b=0}^{B-1} {\mathbf{H}}_{r,k}^{b,b'}(n)}_{\text{ISI across FBMC symbols}}$
\vspace{2pt}
\end{minipage}
\\
\hline

${\mathbf{H}}^{\text{SI},b,b'}_{r,t}(n)$
& Toeplitz residual self loop interference between $t$-th transmit antenna and $r$-th receive antenna at BS. \\
\hline

$\tilde{\mathbf{g}}_{k,t}$ 
& Frequency-selective multi-tap downlink channel vector between the $t$-th transmit antenna of the BS and the $k$-th user. \\
\hline

$\beta_{k}^{\text{dl}}$ 
& Large-scale fading coefficient associated with the downlink channel of the $k$-th user. \\
\hline

$\tilde{\mathbf{G}}_{k,t}$ 
& Toeplitz matrix constructed from $\tilde{\mathbf{g}}_{k,t}$ representing convolution in the downlink channel. \\
\hline

$\tilde{\mathbf{G}}^{\text{IUI}}_{k',k}$ 
& Toeplitz channel matrix representing inter-user interference (IUI) or user loop interference (ULI) between users $k$ and $k'$. \\
\hline

$\mathbf{y}^{\text{dl}}_{k}(i)$ 
& Received time-domain signal vector at the $k$-th user for the $i$-th FBMC block. \\
\hline

$\mathbf{w}^{\text{dl}}_{k}(i)$ 
& Additive white Gaussian noise vector at the $k$-th user in the downlink. \\
\hline

$\mathbf{r}^{\text{dl}}_{k,b}(n)$ 
& Demodulated received signal at the $k$-th user for subcarrier group $b$ and symbol index $n$. \\
\hline

$\mathbf{G}_{k}^{\text{desire},b}$ 
& Effective desired FBMC-domain channel matrix for the $k$-th user at subcarrier group $b$. \\
\hline

$\mathbf{G}_{k}^{\text{ICI},b}$ 
& Effective inter-carrier interference (ICI) channel matrix for the $k$-th user. \\
\hline

$\mathbf{G}_{k,t}^{b,b'}(n)$ 
& Effective FBMC-domain channel matrix from subcarrier group $b$ to $b'$ with delay index $n$ for the link between the $t$-th transmit antenna and the $k$-th user. \\
\hline

$\mathbf{G}_{k',k}^{\text{IUI},b,b'}(n)$ 
& Effective FBMC-domain channel matrix capturing IUI/ULI between users $k$ and $k'$ across subcarriers and delay taps. \\
\hline

\end{tabular}
\end{table}
\begin{table}[H]
\centering
\footnotesize
\caption{Nomenclature ( SINR and Optimization)}
\label{tab_2}
\vspace{-10pt}
\renewcommand{\arraystretch}{1.1}
\begin{tabular}{ |p{3cm}||p{9.5cm}| }
\hline
\textbf{Notation} & \textbf{Description} \\
\hline

$\boldsymbol{\nu}_{k,b}^{\text{dl,int}}$ 
& Aggregate intrinsic interference at the $k$-th user due to FBMC filtering and symbol overlap. \\
\hline

$\boldsymbol{\nu}_{k,b}^{\text{dl,IUI}}$ 
& Aggregate inter-user interference and loop interference affecting the $k$-th user in the downlink. \\
\hline

$\tilde{\mathbf{w}}_{k,b}^{\text{dl}}$ 
& Effective noise vector after FBMC demodulation and receiver processing. \\
\hline

$\mathbf{g}_{k,b,m}$ 
& Channel frequency response (CFR) vector for the $k$-th user at subcarrier index $m$ and subcarrier group $b$. \\
\hline

$\mathbf{f}_{k,b,m}$ 
& Linear precoding vector for the $k$-th user at subcarrier index $m$ and subcarrier group $b$. \\
\hline

$p^{\text{dl}}_{k,b,m}$ 
& Downlink transmit power allocated to the $k$-th user at subcarrier index $m$ and group $b$. \\
\hline

$p^{\text{ul}}_{k,b,m}$ 
& Uplink transmit power allocated to the $k$-th user at subcarrier index $m$ and group $b$. \\
\hline

$\text{SINR}_{s,k}^{b,m}$ 
& Signal-to-interference-plus-noise ratio of the $k$-th user at subcarrier $(b,m)$ in the downlink/uplink. \\
\hline

$\mathcal{N}_{\text{s},k}^{b,m}, \mathcal{D}_{\text{s},k}^{b,m}$ 
& Desired signal power (numerator) and total interference-plus-noise power (denominator) in the SINR expression. \\
\hline

$\mathcal{H}_{\text{s}}$ 
& Set of all downlink/uplink channel realizations. \\
\hline

$\mathbf{G}_{b,m},\, \mathbf{H}_{b,m}$ 
& Concatenated channel frequency response matrix across all users at subcarrier $(b,m)$ in downlink/uplink. \\
\hline

$R_{\text{s}}^{b,m}$ 
& Achievable downlink/uplink rate at subcarrier $(b,m)$. \\
\hline

$R(\mathbf{p},\mathcal{H})$ 
& Total spectral efficiency of the system for a given power allocation and channel realization. \\
\hline

$\mathbf{p}$ 
& Aggregate power allocation vector comprising both uplink and downlink transmit powers. \\
\hline

$p_{\text{dl}}^{\max}, p_{\text{ul}}^{\max}$ 
& Maximum transmit power constraints at the base station (downlink) and users (uplink), respectively. \\
\hline

$\mathbf{F}_{m}$ 
& Linear transmit precoding matrix at subcarrier index $m$. \\
\hline

$\mathbf{v}_{k,m}$ 
& Linear receive combining vector for the $k$-th user at subcarrier $m$. \\
\hline

$p_{k,m}^{s\,(\tau)}$ 
& Transmit power allocated to the $k$-th user at subcarrier $m$ during iteration $\tau$ of the optimization algorithm. \\
\hline

$\gamma_{s,k}^{m}, \lambda_{s,k}^{m}, y_{s,k}^{m}$ 
& Auxiliary and epigraph variables introduced for reformulating the spectral efficiency maximization problem. \\
\hline

$\mu_{s,k}$ 
& Dual variable associated with the transmit power constraint of the $k$-th user. \\
\hline

$\delta^{(\tau)}, \rho^{(\tau)}$ 
& Step sizes used in iterative updates of the optimization algorithm at iteration $\tau$. \\
\hline

$\big[\mathbf{A}\big]_{m,:}, \big[\mathbf{A}\big]_{:,m}$ 
& $m$-th row and $m$-th column of matrix $\mathbf{A}$, respectively. \\
\hline

$\big[\mathbf{A}\big]_{m,m}$ 
& $m$-th diagonal element of matrix $\mathbf{A}$. \\
\hline

$\big[\mathbf{A}\big]_{:,[m:M:KM]}$ 
& Submatrix of $\mathbf{A}$ formed by selecting columns indexed by $\{m, m+M, \dots, KM\}$. \\
\hline

\end{tabular}
\end{table}

\section{FBMC/QAM Signal Processing Chain}
\label{sec:blockdiag}
The pictorial representation of the FBMC/QAM signal processing chains are provided to clearly illustrate the key operations both in the uplink and downlink.
\subsection{Uplink Processing }
\label{subsec:ul_chain}
\begin{figure}[H]
  \centering
\includegraphics[width=0.88\linewidth]{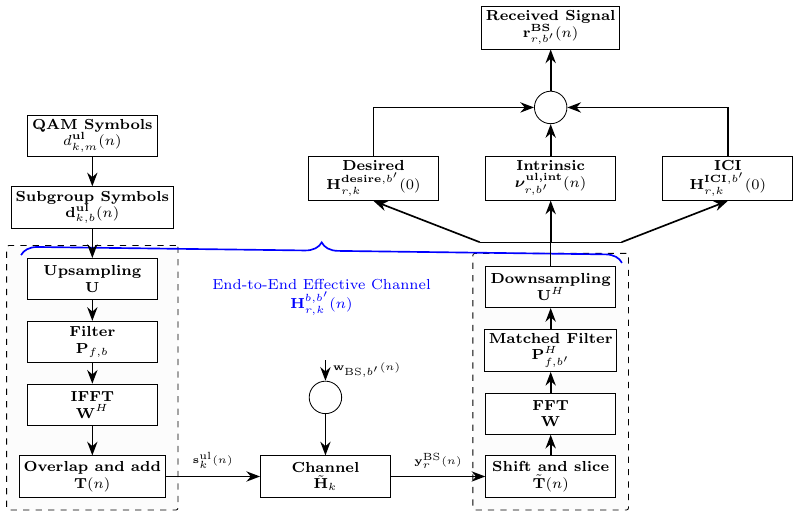}
  \caption{Block diagram of the proposed FBMC/QAM uplink transmission.}
  \label{fig:block_ul}
\end{figure}

The FBMC/QAM uplink processing chain is illustrated in
Figure~\ref{fig:block_ul}.

The signal flow is:
\begin{enumerate}[noitemsep, leftmargin=1.8em]
  \item \textbf{Transmitter (User $k$):}
        QAM symbol vector $\mathbf{d}^{\mathrm{ul}}_{k}$
        $\!\to\!$ subcarrier-group split into $B$ groups $\mathbf{d}^{\mathrm{ul}}_{k,b}$
        $\!\to\!$ upsampling $(\mathbf{U})$ (factor $BL$)
        $\!\to\!$ pulse-shaping filter $\mathbf{p}_b$
        $\!\to\!$ IFFT ($\mathbf{W}^{H}$)
        $\!\to\!$ overlap-and-add $\mathbf{T}(n)$
        $\!\to\!$ transmit $\mathbf{s}^{\mathrm{ul}}_{k}(i)$.

  \item \textbf{Receiver (BS, antenna $r$):}
        Received $\mathbf{y}^{\mathrm{BS}}_{r}(i)$
        $\!\to\!$ shift-and-slice $\tilde{\mathbf{T}}(n)$
        $\!\to\!$ FFT ($\mathbf{W}$)
        $\!\to\!$ matched filter $\mathbf{P}^{H}_{f,b'}$
        $\!\to\!$ downsample ($\mathbf{U}^{H}$)
        $\!\to\!$ $\mathbf{r}^{\mathrm{BS}}_{r,b'}(n')$
        $\!\to\!$ RC vector $\mathbf{v}_{k,b,m}$
        $\!\to\!$ decoded symbol $\hat{d}^{\mathrm{ul}}_{k,m}$.
\end{enumerate}

The received signal $\mathbf{r}^{\mathrm{BS}}_{r,b'}(n')$ decomposes into
(Lemma~2 of main paper):
\begin{equation*}
  \mathbf{r}^{\mathrm{BS}}_{r,b'}(n')
  = \underbrace{\sum_{k}\mathbf{H}^{\mathrm{desire},b'}_{r,k}(0)\mathbf{d}^{\mathrm{ul}}_{k,b'}(n')}_{\text{desired}}
  + \underbrace{\sum_{k}\mathbf{H}^{\mathrm{ICI},b'}_{r,k}(0)\mathbf{d}^{\mathrm{ul}}_{k,b'}(n')}_{\text{ICI}}
  + \boldsymbol{\nu}^{\mathrm{ul,int}}_{r,b'}(n')
  + \boldsymbol{\nu}^{\mathrm{ul,SI}}_{r,b'}(n')
  + \tilde{\mathbf{w}}^{\mathrm{BS}}_{r,b'}(n').
\end{equation*}

\subsection{Downlink Processing}
\label{subsec:dl_chain}

\begin{figure}[H]
  \centering
  \includegraphics[width=0.88\linewidth]{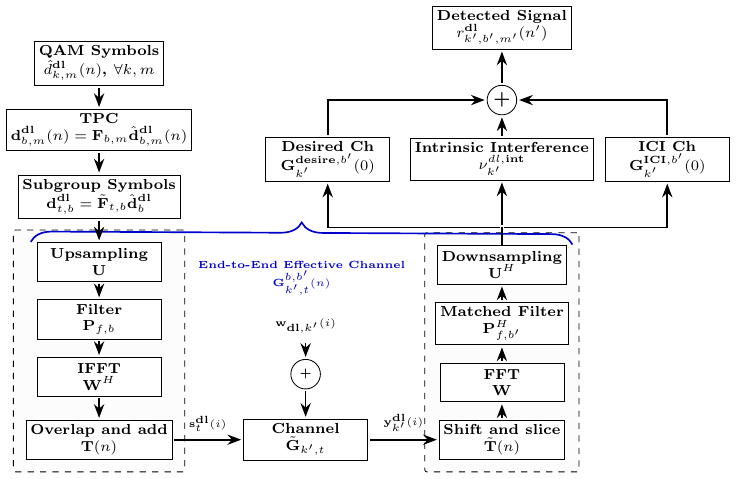}
  \caption{Block diagram of the proposed FBMC/QAM downlink transmission.}
  \label{fig:block_dl}
\end{figure}

The FBMC/QAM downlink processing chain is shown in
Figure~\ref{fig:block_dl}.

The signal flow is:
\begin{enumerate}[noitemsep, leftmargin=1.8em]
  \item \textbf{Transmitter (BS, antenna $t$):}
        $\hat{\mathbf{d}}^{\mathrm{dl}}_{k,m}$
        $\!\to\!$ TPC matrix $\mathbf{F}_{m}$
        $\!\to\!$ subcarrier-group split into $B$ groups ${\mathbf{d}}^{\mathrm{dl}}_{t,b}$
        $\!\to\!$ upsampling $(\mathbf{U})$ (factor $BL$)
        $\!\to\!$ pulse-shaping filter $\mathbf{p}_b$
        $\!\to\!$ IFFT
        $\!\to\!$ overlap-and-add $\mathbf{T}(n)$
        $\!\to\!$ transmit $\mathbf{s}^{\mathrm{dl}}_{t}(i)$.

  \item \textbf{Receiver (User $k'$):}
        $\mathbf{y}^{\mathrm{dl}}_{k'}(i)$ 
        $\!\to\!$ shift-and-slice $\tilde{\mathbf{T}}(n)$
        $\!\to\!$ FFT
        $\!\to\!$ matched filter
        $\!\to\!$ downsample $(\mathbf{U}^{H})$
        $\!\to\!$ $\mathbf{r}^{\mathrm{dl}}_{k',b'}(n')$
        $\!\to\!$ decode $\hat{d}^{\mathrm{dl}}_{k',m}$.
\end{enumerate}
The received signal can be expressed in terms of desired and interference terms (in (30) using Lemma~2) as follow
 \begin{align} \label{eq_ICI,desire}
     {\mathbf{r}}^{\text{dl}}_{k',b'}(n')=\underbrace{{\mathbf{G}}_{k'}^{\text{desire},b'}(0) \tilde{\mathbf{F}}_{b'} {\hat{\mathbf{d}}}^{\text{dl}}_{b'}(n')}_{\text{desired}} +\underbrace{{\mathbf{G}}_{k'}^{\text{ICI},b'}(0) \tilde{\mathbf{F}}_{b'} {\hat{\mathbf{d}}}^{\text{dl}}_{b'}(n')}_{\text{ICI}} 
+ \boldsymbol{\nu}_{k',b'}^{\text{dl},\text{int}}(n') + \boldsymbol{\nu}_{k',b'}^{\text{dl},\text{IUI}}(n')+\tilde{\mathbf{w}}_{k',b'}^{\text{dl}}(n'). \nonumber
\end{align}

\section{Waveform Comparison: OFDM, FBMC/OQAM, FBMC/QAM}
\label{sec:waveform}

The adoption of FBMC/QAM in this work is best understood through the
Balian-Low theorem (BLT), which states that no waveform operating at critical time-frequency (TF) density (TF product $=1$) can simultaneously
achieve all three of:
(i)~perfect complex-field orthogonality,
(ii)~maximum spectral efficiency (no CP or redundancy), and
(iii)~well-localized pulse shapes in both time and frequency~\cite{blt}. Each practical waveform relaxes one condition.
Table~3 summarizes how CP-OFDM, FBMC/OQAM, and FBMC/QAM
navigate this fundamental trade-off.
\begin{table}[H]
\centering
\caption{Comparison of OFDM, FBMC/OQAM and FBMC/QAM.}
\footnotesize
\renewcommand{\arraystretch}{1.2}
\begin{tabular}{|l|r|r|r|}
\hline
\textbf{BLT Property} & \textbf{OFDM} & \textbf{FBMC/OQAM} & \textbf{FBMC/QAM} \\
\hline
Orthogonality 
&  Complex-field 
&  Real-field only 
&  Complex-field \\
\hline
Time-localization 
& Good (rectangular pulse) 
& Good (well-shaped pulse) 
&  Good \\
\hline
Freq-localization 
&  Poor (sinc-shaped) 
&  Excellent 
&  Poor \\
\hline
Max. Symbol Density 
&  1 complex symbol / unit 
&  1 real symbol / 0.5 unit 
&  1 complex symbol / unit \\
\hline
Cyclic Prefix 
&  Required 
&  Not required 
&  Not required \\
\hline
\end{tabular}

\end{table}
From Table~3, it is clear that both OFDM and FBMC/QAM satisfy bi-orthogonality and maximum symbol density but suffer from poor frequency localization. In contrast, FBMC/OQAM achieves good time–frequency localization and maximum symbol density by relaxing orthogonality to the real domain.

\section{Choice of Prototype Filter}
This section presents a comprehensive description of the prototype filter selection employed in the proposed FBMC/QAM system.  It addresses three questions that
arise naturally from the main paper: (i)~\emph{Why is the PHYDYAS filter used?} (ii)~\emph{Is there an
optimal prototype filter for FBMC/QAM?} and (iii)~\emph{How does PHYDYAS
compare against alternative filters from the literature?}
This section provides the rationale for the adopted prototype filter and positions it relative to alternative designs reported in the literature.

\subsection{Role of number of subcarrier groups $B$}
\label{subsec:subcarrier_grouping}
The FBMC/QAM system in the main paper groups the $M$ subcarriers into $B$
groups and assigns a distinct pulse-shaping filter to each group i.e., $\{\mathbf{p}_{\text{tx},0}, \mathbf{p}_{\text{tx},1}, \cdots, \mathbf{p}_{\text{tx},B-1}\}$. This
generalized formulation ($B \geq 2$) is motivated by the need to suppress
intrinsic interference: since the dominant interference in FBMC systems arises
from adjacent subcarriers, assigning different filters to neighbouring groups
can significantly reduce ICI.

In practice, $B=2$ is sufficient and universally adopted~\cite{Base_tcom,nam2014new,freq_2,Equalization},
corresponding to even- and odd-indexed subcarrier groups with two
\emph{sibling} filters.  The justification is:
\begin{itemize}[noitemsep, leftmargin=1.8em]
  \item Non-adjacent subcarriers contribute negligible interference due to the
        good TF localisation of the prototype filter.
  \item Designing $B>2$ mutually orthogonal filters for FBMC/QAM is an open
        and challenging problem with no widely accepted solution in the
        literature.
  \item The PHYDYAS project~\cite{Bellanger2010Primer} provides a well-established
        dual-filter ($B=2$) structure based on block-interleaved orthogonality,
        which is directly used in this work.
\end{itemize}

\textit{Remark:} All simulations in the main paper use $B=2$ with even-subcarrier (base) and
odd-subcarrier (sibling) PHYDYAS filters.  The general $B$ subcarrier groups formulation in
Section~II is retained for mathematical completeness and to accommodate future
extensions.

\subsection{Orthogonality Conditions and the Fundamental Design Constraint}
For interference-free reconstruction under an AWGN channel, the prototype
filters must satisfy the following approximate orthogonality conditions
(derived in Appendix~B of the main paper):
\begin{align}
\mathbf{P}_{f,b'}^{H}\mathbf{W}\mathbf{T}(n)\mathbf{W}^{H}\mathbf{P}_{f,b}
  &\approx \mathbf{0}_{LM}, \quad \forall b,\; n\neq 0, \nonumber \\
\mathbf{P}_{f,b'}^{H}\mathbf{W}\mathbf{T}(0)\mathbf{W}^{H}\mathbf{P}_{f,b}
  &\approx \mathbf{0}_{LM}, \quad b\neq b',         \nonumber      \\
\mathbf{P}_{f,b'}^{H}\mathbf{W}\mathbf{T}(0)\mathbf{W}^{H}\mathbf{P}_{f,b}
  &\approx \mathbf{I}_{LM}, \quad b=b'.                   \nonumber
\end{align}

Under the AWGN assumption and using PHYDYAS-based sibling filters, these
conditions are satisfied to a close approximation. However, in practical frequency-selective fading channels, the conditions
become channel-dependent and are no longer satisfied regardless of the
filter choice:
\begin{align}
\mathbf{P}_{f,b'}^{H}\mathbf{W}\tilde{\mathbf{T}}(n)\tilde{\mathbf{G}}_{k',t}
  \mathbf{W}^{H}\mathbf{P}_{f,b}
  &\neq \mathbf{0}_{LM}, \quad \forall b,\; n\neq 0, \nonumber \\
\mathbf{P}_{f,b'}^{H}\mathbf{W}\tilde{\mathbf{T}}(0)\tilde{\mathbf{G}}_{k',t}
  \mathbf{W}^{H}\mathbf{P}_{f,b}
  &\neq \mathbf{0}_{LM}, \quad b\neq b', \nonumber \\
\mathbf{P}_{f,b'}^{H}\mathbf{W}\tilde{\mathbf{T}}(0)\tilde{\mathbf{G}}_{k',t}
  \mathbf{W}^{H}\mathbf{P}_{f,b}
  &\neq \mathbf{I}_{LM}, \quad b=b'. \nonumber
\end{align}

These properties are generally not satisfied unless the prototype filters are carefully designed by accounting for channel effects. Furthermore, when employing separate filters across subcarrier groups (e.g., even and odd groups, or more generally $B$ groups), the absence of strict orthogonality further aggravates ICI. \\

To validate the above statement, we have presented a bar chart, demonstrating the breakdown of signal (desired $+$ interference) components for the proposed multi-user FBMC/QAM system associated with a PHYDYAS filter for a downlink scenario as follows  

\begin{center}
\includegraphics[width=0.7\linewidth]{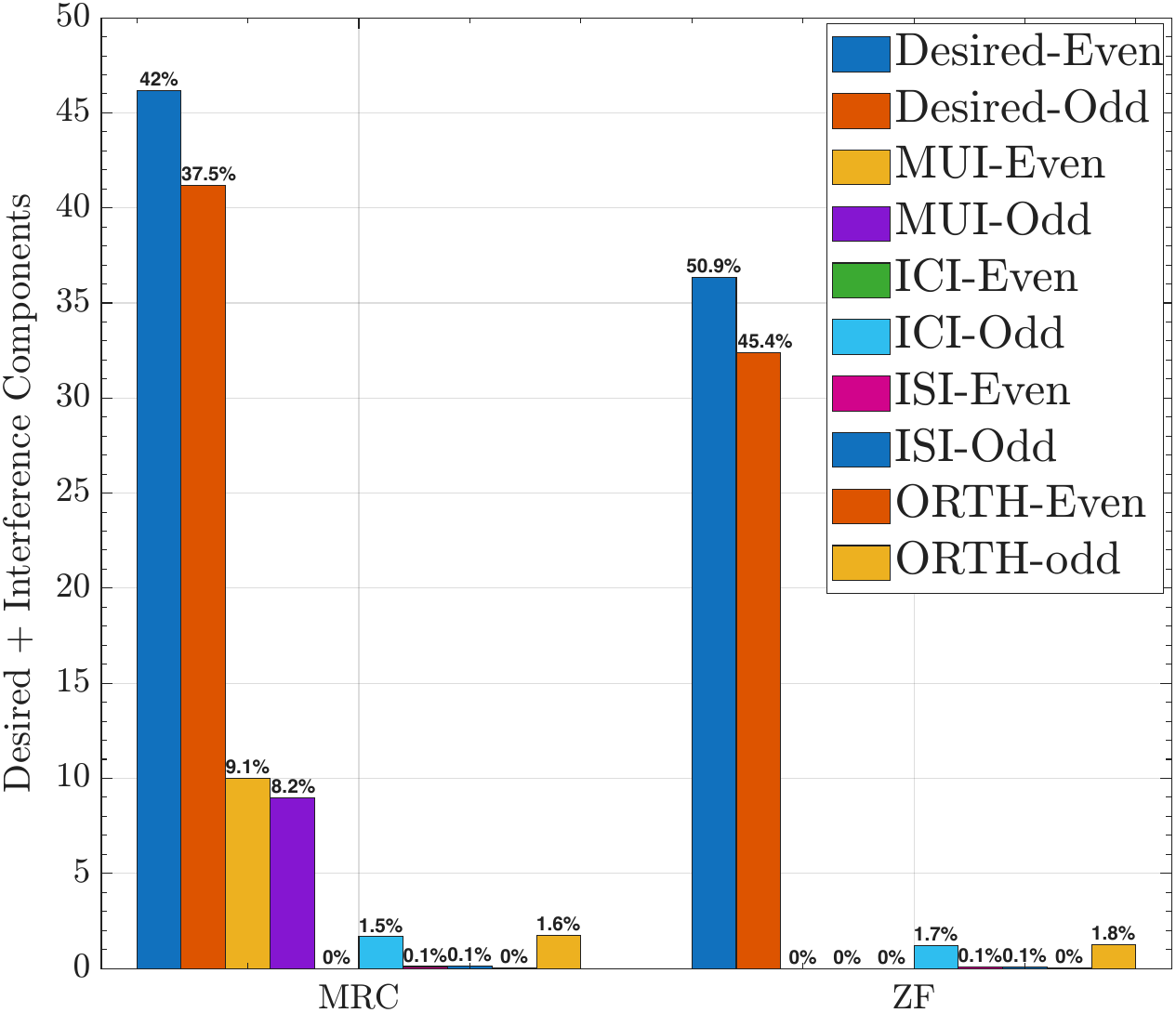}\vspace{-5pt}
  \captionof{figure}{
Desired and interference strength breakdown for MRT and ZF TPC schemes.}
  \label{fig:fig4_1}
\end{center}
Figure~3 shows the desired signal and interference breakdown of the proposed FBMC/QAM system by plotting a bar chart for different transmit precoding (TPC) schemes for a fixed transmit power of $P_{t}$ $=15$ dB. For this simulation, the number of users and transmit antenna (TA) are set as $K=8$ and $N_{\text{tx}}=32$. The legend entries "Desired-Even" and "Desired-Odd" refer to the desired signal strength for even-and odd-subcarrier symbols. Likewise, \{"MUI-Even","MUI-Odd"\}, \{“ICI-Even”, “ICI-Odd”\}, \{“ISI-Even”, “ISI-Odd”\} and \{“ORTH-Even”, “ORTH-Odd”\} represent the multi-user interference (MUI), ICI, ISI and the orthogonality lost among even and odd subcarrier symbols due to violation of orthogonality conditions in (2), respectively. It is observed that MUI is the dominant interference under the MRT scheme, while for ZF, the major contributors are ICI-Odd and ORTH-Odd. This is attributed to the fact that the proposed FBMC/QAM scheme exploits orthogonal filters for the even-subcarrier symbols and nearly-orthogonal filters for the odd-subcarrier symbols. For the same reason, it is also observed that the desired signal strength for the even indexed subcarrier symbols is higher for both the MRT (42\%) and ZF (50.9\%) schemes compared to the odd subcarrier symbols. By contrast, interference components including ICI, ISI, and loss of orthogonality (ORTH) are more dominant in the odd subcarrier symbols.

\subsection{Filters Comparison: PHYDYAS, Type-I, and Type-II}
\label{subsec:filter_list}

To demonstrate that the proposed framework is not limited to the PHYDYAS filter, we additionally evaluate its performance using the Type-I and Type-II prototype filters proposed by Yun et al. \cite{yun2015new}.

\label{subsec:filter_results}

\begin{figure}[H]
\centering
\includegraphics[width=0.6\linewidth]{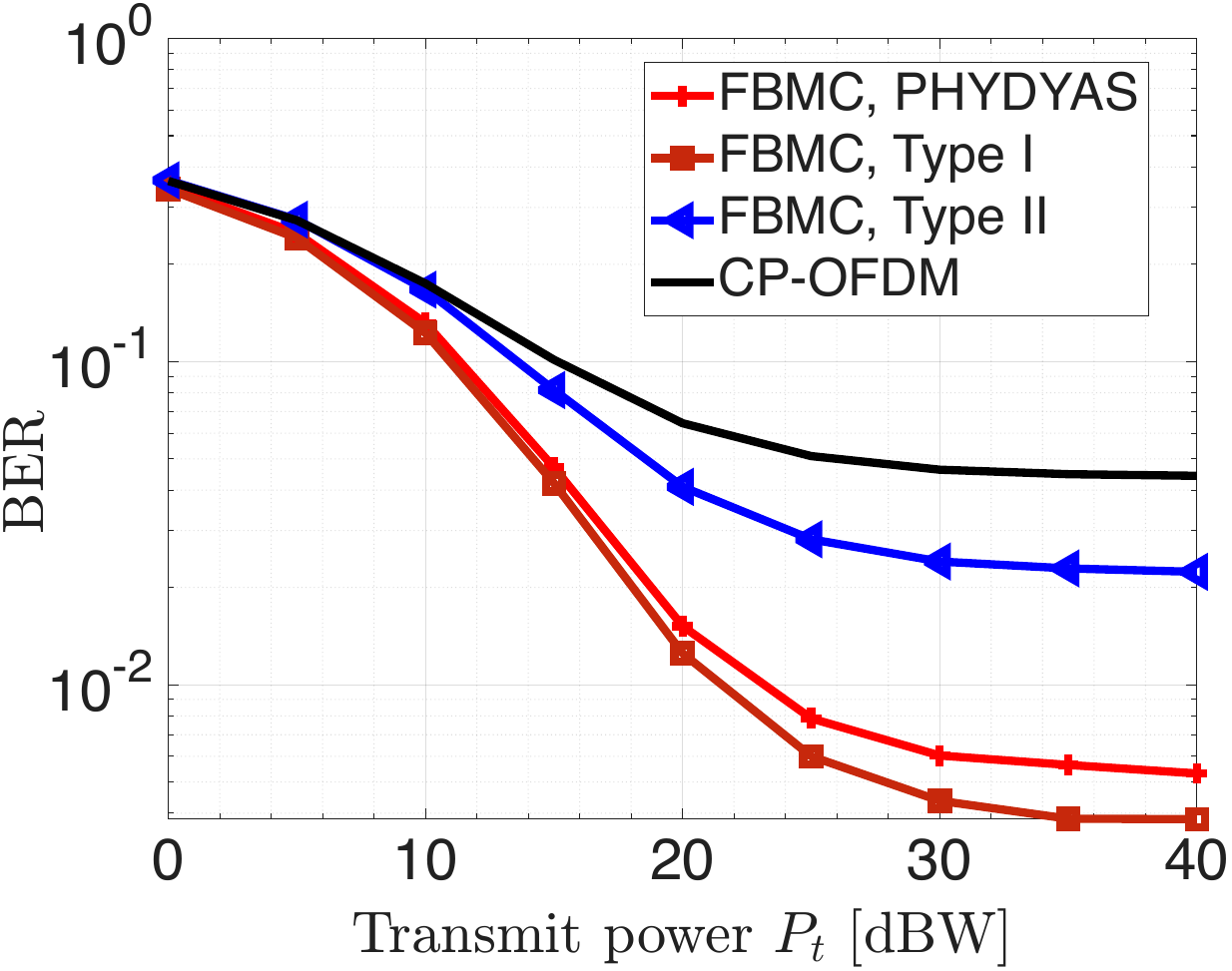}
\caption{BER versus transmit power for the proposed FBMC/QAM system using (i) PHYDYAS, (ii) Type-I, and (iii) Type-II prototype filters.}
\label{fig:fig2_3}
\end{figure}

\begin{figure}[H]
\centering
\includegraphics[width=0.6\linewidth]{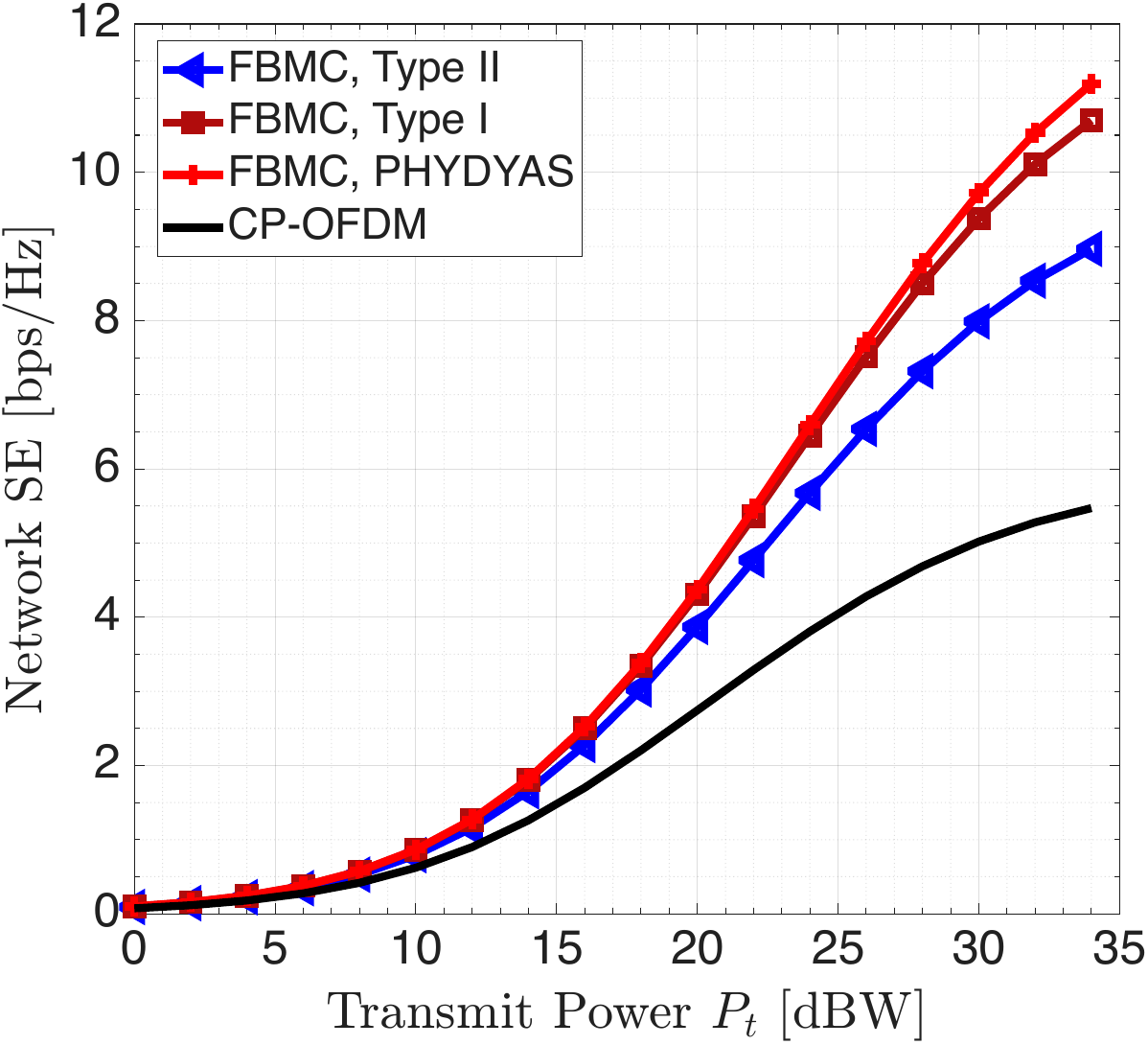}
\caption{Network SE versus transmit power for the proposed FBMC/QAM system using (i) PHYDYAS, (ii) Type-I, and (iii) Type-II prototype filters.}
\label{fig:fig2_4}
\end{figure}

Figures~\ref{fig:fig2_3} and~\ref{fig:fig2_4} compare BER and
network SE for all three prototype filters under a $K=2$ user setup with
$N_{\mathrm{tx}}=N_{\mathrm{rx}}=8$, perfect SI cancellation, and residual
CFO $\varepsilon=0.3$. 

Figure~\ref{fig:fig2_3} shows the BER versus transmit power of the proposed FBMC/QAM system for different prototype filters under a $K = 2$ user setup. For comparison, the CP-OFDM counterpart is also considered. In addition, perfect SI cancellation is assumed in this setup. It is observed that the proposed FBMC/QAM system achieves superior BER performance compared to CP-OFDM under residual CFO (with $\epsilon = 0.3$) for all the considered prototype filters. Furthermore, consistent with the observations in Fig.~3 and Fig.~4 of~\cite{freq_1}, the Type-I filter exhibits better BER performance compared to the Type-II and PHYDYAS filters.
Figure~\ref{fig:fig2_4} presents the FD network SE performance of the proposed FBMC/QAM system in comparison with CP-OFDM under CFO impairment ($\epsilon = 0.3$). For this setup, the number of users, TAs, and RAs are set to $2$, $8$, and $8$, respectively. In addition, perfect SI cancellation is assumed. It is observed that CP-OFDM suffers from a significant SE degradation compared to FBMC/QAM for all considered prototype filters, highlighting the improved robustness of FBMC/QAM against CFO impairments. Interestingly, unlike the BER results in the Figure~3 where the Type-I filter provides the best performance, the PHYDYAS filter achieves higher network SE compared to the Type-I and Type-II filters. 

\subsection{Recent Advances in Prototype Filter Design for FBMC/QAM}
\label{subsec:filter_advances}
To place the prototype filter selection in the context of recent developments in the literature, the following advances are briefly reviewed.

\subsubsection*{Dual-filter optimization~\cite{Ren2025PrototypeFO}}
Ren~et~al.\ in \cite{Ren2025PrototypeFO} propose a dual-filter structure in which the PHYDYAS filter is
fixed for the even-subcarrier group while the odd-subcarrier filter is
jointly optimized to minimize SI, time dispersion, and frequency dispersion.
Simulation results in~\cite{Ren2025PrototypeFO} demonstrate that the optimized odd-subcarrier
filter improves spectral confinement without degrading BER performance.
This direction is complementary to the system-level design in the main paper
and constitutes a promising avenue for future work.

\subsubsection*{Mismatched transmit-receive filters~\cite{mismatch}}
The study in~\cite{mismatch} proposes a low-complexity FBMC/QAM
framework using mismatched transmit and receive filters (PHYDYAS at the
transmitter, a distinct optimized filter at the receiver), formulated as a
relaxed LASSO problem.  This provides an alternative approach to reducing
self-interference without modifying the transmit-side filter.

\subsubsection*{Towards a unified optimal filter design}
From the BLT perspective (refer to Section~3), any future
optimal filter for FBMC/QAM must navigate the three-way trade-off among
orthogonality, TF localization, and SE.  Currently, no universally accepted
framework for jointly optimizing all three exists in the FBMC/QAM context,
particularly for multi-user MISO/MIMO settings with full-duplex operation.
The design of such a filter, accounting for end-to-end fading channel effects
as provided in Section~4.2 remains
an open research problem.
\section{Comparison of the proposed online stochastic optimization with the offline LDT-based optimization in \cite{QT_1}}

The proposed scheme aims to maximize the ergodic sum rate (refer to the $\mathbf{P}1$ in Section~III-A)
\begin{align}
R_{\mathcal{S}}(\mathbf{p}) \triangleq \mathbb{E}_{\mathcal{H}}\big[ R(\mathbf{p}, \mathcal{H}) \big], \nonumber
\end{align}
subject to the given uplink and downlink power constraints. Since the expectation over the channel realizations is not available in closed-form, we adopt a stochastic approximation approach.
Specifically, at the $\tau$-th iteration, the objective function is replaced by its stochastic estimate $\tilde{R}^{(\tau)}_{\mathcal{S}}(\mathbf{p})$, updated as
\begin{align}
\tilde{R}^{(\tau)}_{\mathcal{S}}(\mathbf{p})
= (1-\delta^{(\tau)}) \tilde{R}^{(\tau-1)}_{\mathcal{S}}(\mathbf{p})
+ \delta^{(\tau)} \hat{R}(\mathbf{p}; \mathbf{p}^{(\tau)}, \mathcal{H}^{(\tau)}), \nonumber
\end{align}
where $\mathcal{H}^{(\tau)}$ denotes the channel realization observed at iteration $\tau$.
By recursively expanding the above expression, we obtain the following expression.
\begin{align}
\tilde{R}^{(\tau)}_{\mathcal{S}}(\mathbf{p})
= \sum_{i=0}^{\tau} w_i^{(\tau)} \, \hat{R}(\mathbf{p}; \mathbf{p}^{(i)}, \mathcal{H}^{(i)}),\nonumber
\end{align}
where the weights are given by
\begin{align}
w_i^{(\tau)} =
\begin{cases}
\displaystyle \prod_{j=1}^{\tau} (1-\delta^{(j)}), & i = 0, \\[2ex]
\displaystyle \delta^{(i)} \prod_{j=i+1}^{\tau} (1-\delta^{(j)}), & i = 1,\dots,\tau,
\end{cases}
\nonumber\end{align}
and satisfy $w_i^{(\tau)} \geq 0$ and $\sum_{i=0}^{\tau} w_i^{(\tau)} = 1$. Hence, $\tilde{R}^{(\tau)}_{\mathcal{S}}(\mathbf{p})$ is a convex combination (i.e., weighted average) of past instantaneous rate samples. In particular, when $\delta^{(\tau)} = 1/(\tau+1)$, the above reduces to the sample average over $\tau+1$ channel realizations. For instance, starting from the initialization $\tilde{R}^{(0)}_{\mathcal{S}}(\mathbf{p}) = \hat{R}(\mathbf{p}; \mathbf{p}^{(0)}, \mathcal{H}^{(0)})$, and applying induction, $\tilde{R}^{(\tau)}_{\mathcal{S}}(\mathbf{p})$ reduces to
\begin{align}
\tilde{R}^{(\tau)}_{\mathcal{S}}(\mathbf{p})
= \frac{1}{\tau+1} \sum_{i=0}^{\tau}
\hat{R}(\mathbf{p}; \mathbf{p}^{(i)}, \mathcal{H}^{(i)}).\nonumber
\end{align}
Thus, $\tilde{R}^{(\tau)}_{\mathcal{S}}(\mathbf{p})$ is exactly the arithmetic mean of the instantaneous rate samples obtained from the first $\tau+1$ channel realizations. 
This demonstrates that the proposed method performs online averaging of the ergodic objective, using one channel realization per iteration. Consequently, after $\tau$ iterations, the method effectively utilizes $\tau+1$ snapshots, without requiring storage of past channel realizations. The stopping criterion $\tilde{R}^{(\tau)}_{\mathcal{S}}(\mathbf{p})-\tilde{R}^{(\tau-1)}_{\mathcal{S}}(\mathbf{p})\leq \epsilon$ confirming the convergence of the SSCA algorithm. As verified numerically in Fig.~3 of the main manuscript, convergence is achieved within 100 iterations.

In contrast, the method in \cite{QT_1} approximates the ergodic objective using a batch Monte Carlo approach:
\begin{align}
R_{\mathcal{S}}(\mathbf{p})
\approx \frac{1}{N} \sum_{i=1}^{N} R(\mathbf{p}, \mathcal{H}^{(i)}), \nonumber
\end{align}
%and solves
%\begin{align}
%\mathbf{p}^\ast = \arg\max_{\mathbf{p}} \frac{1}{N} \sum_{i=1}^{N} R(\mathbf{p}, \mathcal{H}^{(i)}),
%\end{align}
which requires access to all $N$ channel realizations simultaneously. since the offline 
method approximates the ergodic objective via sample averaging, i.e., 
$\mathbb{E}[R(\mathbf{p},\mathcal{H})] \approx \frac{1}{N}\sum_{i=1}^{N} 
R(\mathbf{p},\mathcal{H}^{(i)})$, the required number of snapshots $N$ to 
ensure convergence of the sample mean to the true expectation is not known 
a priori. 
Based on the above discussion, Table 4 has been added to compare the proposed algorithm with the LDT-based approach in \cite{QT_1}.
\begin{table}[H]
\caption{Comparison of the Proposed Algorithm and LDT \cite{QT_1}.}
\label{tab:complexity}
\centering
\small
\begin{tabular}{|l|r|r|}
\hline
\textbf{Aspect} &
\textbf{Algorithm 1} &
\textbf{LDT \cite{QT_1}} \\
\hline
Monte Carlo averaging & \textbf{No} & Yes \\
\hline
External solver (CVX) & No & No \\
\hline
Power update & Closed form & Closed form \\
\hline
SE performance & Comparable & Benchmark \\
\hline
Convergence requirement & \textbf{Small iterations} & Large iterations \\
\hline
\end{tabular}
\end{table}

%Therefore, $N$ is set sufficiently large ($N = 500$) to guarantee 
%statistical reliability in accordance with the law of large numbers.

%Note that , the parameter $N$ need to be large enough to estimate the expectation $\mathbb{E}_{\mathcal{H}}\big[ R(\mathbf{p}, \mathcal{H}) \big]\approx \frac{1}{N} \sum_{i=1}^{N} R(\mathbf{p}, \mathcal{H}^{(i)})$. For the simulations in Fig.~2c, $N=500$ snapshots are used in [30] to ensure accurate estimation of the expectation.

\bibliographystyle{IEEEtran}
%\bstctlcite{IEEEexample:BSTcontrol} 
\bibliography{biblio}
\end{document}